# INFORMATION ASYMMETRY INDEX: The View of Market Analysts

**Roberto Frota Decourt**
**UNISINOS – Universidade do Vale do Rio dos Sinos**

**Heitor Almeida**
**University of Illinois at Urbana–Champaign**

**Philippe Protin**
**Université Grenoble Alpes**

**Matheus R. C. Gonzalez**
**UNISINOS – Universidade do Vale do Rio dos Sinos**

**Abstract:** The purpose of the research was to build an index of informational asymmetry with market and firm proxies that reflect the analysts' perception of the level of informational asymmetry of companies. The proposed method consists of the construction of an algorithm based on the Elo rating and captures the perception of the analyst that choose, between two firms, the one they consider to have better information. After we have the informational asymmetry index, we run a regression model with our rating as dependent variable and proxies used by the literature as the independent variable to have a model that can be used for other researches that need to measure the level of informational asymmetry of a company. Our model presented a good fit between our index and the proxies used to measure informational asymmetry and we find four significant variables: coverage, volatility, Tobin q, and size.

## 1 INTRODUCTION

Information asymmetry can be perceived over different forms, many researches have been dedicated to study each of these different impacts on companies. Verrechia (1983), Myers and Majluf (1985), Easley and O'Hara (1987), Harris and Raviv (1991) and Wang (1993) show theoretically that opaquer companies tend to have more noise, volatility and uniformed traders acting on their stock price, as well more conflict between managerial and shareholders which impacting on firm's investment decision and capital cost. Under those circumstances, information asymmetry has been receiving much attention of modern literature.

As seen in Beyer et al (2010) many academics have been developing their researches using different proxies, quantitative and qualitative variables, looking to verify the impact of information asymmetry into the companies in terms of executive compensation, cost of capital, level of indebtedness, the company's profitability, shareholder return, liquidity, control structure and dividend policy, but so far haven't found appropriate proxies to measure the information asymmetry degree. Most corporate finance literature have been using proxies from financial analysts forecast, company's investment opportunities, and the presence of informed and uniformed traders over daily stock prices.

This work conducted a survey with financial analyst in order to verify their perception of disclosure and information asymmetry over companies which belong to Brazil Broad-Based Index (IBrA). Financial analyst was invited to access a web site an choose among a pair of companies which one have better disclosure. The analysts must be certified by Analysts and Investment Professionals of Capital Markets Association (*APIMEC*), or Charter Financial Analysts (CFA) holder

Based on Elo (1961) algorithm this work creates the Information Asymmetry Index (IAI) classifying Brazilian companies' disclosure under market perception. The logic of this ranking is to check the likelihood of a win (lost) between direct disputes. A win when the expected probability was high would add very few points to the ranking, however, a win with a very small probability adds many points to the ranking. On the other side, a lost with high probability lose few points and, a lost with low probability lose many points.

A concern of this work is if the IAI correctly capture the disclosure perception of financial analyst, or it is disturbing by other sources besides disclosure. In order to test the robustness of IAI correct, three tests were conducted base on literature review of the theme. The tests are divided into three categories, separated accordingly proxies' sources, named: external analysis, internal analysis and market microstructure. First and second group used panel data regression and the third uses likelihood function.

External analysis is a group characterized by information that came from outside de company, i.e. financial analysts. They are very qualifying professional who work to understand every companies' detail for the purpose of determine what should be the stock price in a specific period. Analyst coverage, analyst forecast dispersion and the absolute difference between the actual earnings and the Bloomberg median forecast were the proxies used to measure information asymmetry in this group.

Internal analysis looks for information asymmetry trough companies financial report and stock price volatility. Shin and Stulz (2000) studied the relation between Tobins'q and systematic equity risk and total equity risk. They find that firms with higher market to book ratio have higher growth opportunities. Also, because firms' value is composed by the value of real asset plus growth opportunities, they suggested that firms with higher growth opportunities have more volatility. Literature suggests appropriate proxies of information asymmetry under companies' financial statement and stock price. This work uses growth opportunities proxy, Research and Development and stock price volatility as measures of information asymmetry in this group.

Market microstructure investigated the likelihood of informed investors determined stock price. Easley, Kiefer and O'Hara (1997), Easley, Hvidkjaer, and O'Hara (2002), Easley and O'Hara (2004) and (Hvidkjaer, 2010) develop a model to demonstrate that informed investors not only plays a critical component on stock price formation but also build a different portfolio from other investors, which is the opposite that CAPM theory sustained. This work verifies the probability of informed trading in the same days that the survey was conducted trying to understand if companies with higher probability of informed investors presence are associated with low levels of IAI.

The finance theory's premise that companies and their management work to maximize value to their stockholders, i.e. get the company value as high as possible. In this sense emerges from the corporate governance the best practices that could lead the company on the way of transparency and disclosure actions, achieving a lower degree of risk, so market starts to trust the company and their executives. Hence, companies should strive to a higher level of disclosure, even though would be an additional cost involved in it.

In effort to measure the impact of information asymmetry on companies, several studies have been conducted, because the level of information asymmetry between company managers and the market may have different consequences for each company, such as executive compensation, cost of capital, level of indebtedness, profitability, shareholder return, liquidity, governance and dividend policy, consequently managers are constantly in a trade-off about what information will be disclosed by the companies (Beyer, Cohen, Lys & Beverly. 2010).

Because of information asymmetry has a large impact on companies, there are plenty methodologies to evaluate this construct likewise disclosure and quality reports (Beyer et al, 2010). Researchers bias their analyzes by perceiving information asymmetry only through the eyes of their work, but not in the all impacts it may have on firms. Academics who observes information asymmetry in initial public offering usually use growth opportunities as a proxy, living aside important details like how competitive the market is, how many hours of meeting did the company had before goes public and if the company already has stocks on other stock exchange, all off it, and more, may impact increase (decrease) the information asymmetry od the entrance company. Another strand of researcher limits its construct on firm's expected future earnings and market forecast, although it might be a relevant proxy, there isn't relevant news every day, while companies' stock price can have large volatility even in the days with absence news. For the same reason the presence of uninformed traders can dramatically change over time, not only by companies' news, but by changes in economy, survivor ship bias and other behavioral issues and size of the market diminishing.  All things considered, the absence of a methodology capable to capture all the aspects of information asymmetry reduces external validity of researchers. That is the central concern of this work, provide a wide and highly accurate method to measure information asymmetry.

The purpose of this research is to create a raking of Brazilian companies' disclosure through market participant perspective. The ranking will be provided by the Information Asymmetry Index which must be highly associated and statically significant correlated with usual methods of measure information asymmetry provided by the literature.

As seen in Beyer et al (2010) many academics have been developing their researches using different proxies, quantitative and qualitative variables, looking to verify the impact of information asymmetry into the companies in terms of executive compensation, cost of capital, level of indebtedness, the company's profitability, shareholder return, liquidity, control structure and dividend policy, but until now, hasn't been found an effective way to measure it.

Although there are plenty researches over information asymmetry, the variability of the methodology used on then contributes negatively to the literature in a view of it decreases the research's external validity. To my knowledge, few researches attempt build a wide methodology to cover all aspects of information asymmetry and no one is based on market participant perception adherent on literature. Thus, this work aims to creates an index under financial analyst perception and test it over and test it with what the literature indicates as proxies of informational asymmetry,

Investment analysts have extensive work to do, to determine companies' value they need to fully understand its business, read their financial statement (including footnotes and some accessories commentary). Moreover, they are influenced by the cost of achieving information and most important their capacity to prove their right instead of the market (Brennan & Tomarowsky, 2000). Hence, investment analysts are too deep in companies' day by day, figures and disclosures practices. In additional investment analysts are concerned about liquidity which measures the investor's demand of a stock, Kyle (1985) said that as higher is the number of shares traded, less would be the degree of asymmetry information. In other words, Bushe and Miller (2012) stated that firms with low visibility and poor disclosure programs move away from security analysts and institutional investors. Hence, companies with disclosure policy can enhance their liquidity (Botosan, 1997) which can lead the market to a more accurate pricing. Since investment analysts contribute to enhancing capital markets by through their corporate reports, valuations and forecasts (Healy & Palepu, 2001), is plausible to assume that they are one of the most qualify agents to evaluate companies' disclosure, which consequently can decrease asymmetry information (Diamond & Verrecchia, 1991). This work chooses to survey the opinion of certified analyst by CFA institute or APIMEC, besides a good analyst could not hold one of it titles, the ones who hold it certainly posses the knowledge to conduct a great valuation and interpretation of companies' figures. Besides that, both institutes helped the research by ask their affiliates to answer the survey.

## 2. LITERATURE REVIEW

An important asset in finance and economy is information, Stiglitz (2017) stated that about a century ago economists started to study information economics, developing models carrying out the presumption of market efficiency aiming to understand the economic police impacts. These studies revealed that quite often markets aren't efficiency, consequently information plays an important driver in the efficient capital allocation.

Information and knowledge are substantially different from ordinary good studied by economists, due it's global and public characteristics (Stiglitz, 1995). Verrechia (1983) defined it as "is a signal which reveals the true liquidating value of the risky asset perturbed by some noise" (p. 179) and Usategui (2002) complemented this definition by adding that "The most valuable information is the one that solves the uncertainty of the decision maker" (p. 136). The value of information is a puzzle to complete, Usategui (2002) argued that an information provides the decision maker a higher expected return, consequently, the value of an information is the difference between expected return with an additional information vis-à-vis the expected return without it. In his words, agents would be willing to pay this entire difference. Even though it's plausible method of evaluating information, doesn't seem coherent that agents are willing to pay all the extra earning for that information. Paying it they would be incurred some additional costs, turning the expected profit at the same level of was before without possessing the information. A fraction of the extra earing would be more reasonable, but still facing the problem of which fraction would be fair.

## 2.1 Information Asymmetry

Fama (1969) said that an efficient market is the one which security prices at all the time “fully reflects” plenty available information. In fact, this term is so general, that makes it difficult to test it empirically. Initially, he states some conditions to market efficient: information is costless, available to all and easily understandable. In brief there are three empirical teste categories depends of the information interest: weak, semi-strong and strong form. In the weak form was tested if the information interest was historical price, he found evidences that daily price changing were dependent proving a serial correlation, but close to zero, in additional, Fama said that an overreaction, might be followed by a large price changes, although with unpredictable sign, showing that investors take a while to understand and evaluate the new information, even though he found that the first day’s announcement is unbiased. Testing semi-strong form is a format that stock prices fully reflected all public information supports the theory of efficient market, i.e. future dividend payments, split announcements, earnings announcements, new issues, or other information are on average fully reflected in the prices. And also, there is the strong form which prices reflect all available information, however, two important deviations had been found that some highly influenced market agents have access to information before than others, making profit with it, and some corporate insiders can have access to some exclusive information about their companies, but even their price deviation would permanently persist.

Although, many researchers have been criticizing this view, Brennan and Tamarowski (2000) say that the initial conditions for market efficiency are strongly wrong in practice. Managing a company is truly complex, they must be aware of external threats, internal conflicts and they often sell technically sophisticated products, which may impact on share’s values and can lead financial market to misprice it.

On the other hand, the studying of market efficiency and information asymmetry have been emerging in areas like accounting, finance, and corporate governance. Akerlof (1970), describe the market of lemons, where informal guarantees and asymmetric information take place, in other words, adding the construct “trust” into an economic model. He noticed that in the market of used cars in America, asymmetry information was inherent. Because buyers can’t identify the difference between good and bad cars, which are traded at the same price, the sellers of good cars would be discouraged to offer their assets, since they wouldn't get the expected value for the car, but in fact, the value of a lemon car. This process named Adverse Selection were detected in other markets too. Usategui (2002) examined this practicing between companies and banks, stating that companies might have their own resources needed to finance a project, although as they are risk-averse, they're going to take a loan in a bank which is risk averse too, but in a lower rate. Whether the bank knows the risk distribution of his credit portfolio, the interest rate charged in each project would be some that represent the average risk of all the loans. Hence, companies with lower risks may finance their projects with internal funds, for this reason, banks would have creditors with higher risks. As in the market of lemons (Akerlof, 1970) by the adverse selection, only companies with high credit risks are going get a bank loan, turning market worse. So, might be plausible to assume that this phenomenon can take place when a company goes public. Underwriters force managements to issue equities below their expected return (Stoll & Curley, 1970), if it is truth, only companies which doesn't have internal funds would go public, turning market poorer, putting away good companies.

Because information is difficult to evaluate, in the context of corporate finance, literature has brought the notion the firm's insiders are well informed than market participants, some researcher have been dedicating to study what is called "conflict of interest" especially in the relations between equity holders and managers and between equity holders and debtholders.

Harris and Raviv (1991) in their seminal article, provided a review of what had been written so far about agency costs, asymmetric information, and other topics. Agency costs is the cost due to conflict of interest, Harris and Raviv (1991), said it takes place by two different relations: between shareholders and managers and between shareholders and debtholders. The first conflict arises when there isn't an alignment in company corporate governance. Managers, whose don't have shares, can prefer personal compensation and a higher leverage – besides higher profits –, decreasing free cash flow to equity and consequently not maximizing firm value, in this sense managers would be benefits by companies' profit. Consequently, equity holders can be conservative to select companies' investments, even if they have a profitable payout. The second conflict occurs in the relation between debtholders and equity holders, because the covenants contracts lead equity holders to invest sub-optimally, in a process named “asset substitution effect”. Equity holders will capture the gain of an investment only if it yields a return bigger than the cost of debt, otherwise only the debtholders would be benefited. This relation is an incentive to equity holders invest in risky projects, even if they decrease the equity value, aiming to get higher returns. In Brazil, is common a third conflict between minority shareholders and controlling. Rabelo and Vasconcelos (2002) said that ownership is too concentrate, in structure called pyramids, which enhance the power on dominant shareholders, and do not see minority shareholders as partners.

Information Asymmetry also impacts on capital structure and level of indebtedness, (Modigliani and Miller, 1958; Ross, 1977; Myers and Majuf, 1984; Botosan, 1997). Harris and Raviv (1991) stated that internal sources are always preferred than external, to avoiding stock price reaction. However, companies go public to financing, which can implicate in negative reaction on stock price, because investors might conclude that internal sources and riskless debt wasn’t enough, or wasn’t there for the company, requiring higher returns. Moreover, debt issuance is a signaling of asymmetry information. Harris and Raviv (1991) argue that managers are well known about firms’ returns distribution, companies are expected to leverage (deleverage) if current market are lower (higher) than futures. Since investors would expect higher returns if debt level is increasing – as higher quality firms finance issuing more debt and lower quality companies issue more equity to finance – stock prices reaction should be positive in response to debt issuing.

Information Asymmetry has been receiving a relevant attention on the body of corporate finance literature, even though, there isn’t a consensus in how to proxy information risk, since it is not an observable construct, empiricist must rely on proxy variables. Clarke and Shastri (2001) divided in three general classes of proxies:

Internal analysis is the first group, it looks for proxy in order to identify growth opportunities, since companies with higher growth opportunities have a higher degree of information asymmetry Adam and Goyal, 1999; Shin and Stulz, 2000). Literature have been using R&D investments, market-to-book asset ratio and earnings-price,

External analysis is the second group, literature have been using analysts forecast of future earnings as proxy of information asymmetry, researchers find that as long companies increase communication, more accurate stock prices would be, more analyst coverage, less dispersion on analyst forecast and consequently reduction on asymmetry (Lang and Lundholm, 1993; Thomas, 2002, Irani and Karamanou, 2003).

Finally, several papers had payed attention on the adverse selection component of bid-and-ask spread, since market makers are trading with unidentified investors in a competitive environment, they are widening the spread to recover possible losses traded with informed investors (Glosten and Harris, 1988). Literature (Lambert, Leuz and Verrechia, 2008; Armstrong, Core and Taylor, 2011; He, Lepone and Leug, 2013) also examine the relation between information asymmetry and cost of capital and equity (here and after COEC). The findings suggest a positive relation between then, especially when markets are imperfect.

## 3 METHODOLOGY

The Informational Asymmetry Index (here an after IAI) was created to capture the analysts' perceptions about the level of company's disclosure and information asymmetry. IAI is based on Elo ratings, which was developed by Arpad Elo (1961) and is best known as the ranking system used to rank chess players.

The logic of this ranking is to check the likelihood of a win (lost) between direct disputes. A win when the expected probability was high would add very few points to the ranking, however, a win with a very small probability adds many points to the ranking. On the other side, a lost with high probability lose few points and, a lost with low probability lose many points.

The IAI will use this method on all pair company dispute which were answered by market analysts accredited on APIMEC (Analysts and Investment Professionals of Capital Markets Association) or CFA holder (Chartered Financial Analyst). Hence the IAI was able to capture the disclosure of a large number of Brazilian companies from the market perspective.

In order to exemplify this logic, let's assume a dispute among two companies, company X (Elo-rating score: 1,200) and company Y (Elo-rating score: 1,000). The difference between rankings is 200 points, which would represent a win probability of 76% for X and 24% for Y, according table 5 presented by Albers and Vries (2001). The new companies' score would be as follow:

Equation 1: IAI score

$$New\ X\ Score = Previous\ Rank\ Score + (1 - p)k$$

$$New\ X\ Score = [1200 + (1 - 0{,}76)] \times 100$$

$$New\ X\ score = 1224$$

$$New\ Y\ Score = Previous\ Rank\ Score - (Score\ added\ by\ the\ winner)$$

$$New\ Y\ Score = 1000 - 24\ = 976$$

Source: The author

In this example, $p$ is the win probability and $k$ is a constant, which will be discussed later.

Table 5: Difference in Elo-rating and the corresponding win expectation

| Rating difference | Expected chance of winning | Difference | Chance | Difference | Chance |
|---|---|---|---|---|---|
| 0>=dif<=3 | 0.50 | 122>=dif<=129 | 0.67 | 279>=dif<=290 | 0.84 |
| 4>=dif<=10 | 0.51 | 130>=dif<=137 | 0.68 | 291>=dif<=302 | 0.85 |
| 11>=dif<=17 | 0.52 | 138>=dif<=145 | 0.69 | 303>=dif<=315 | 0.86 |
| 18>=dif<=25 | 0.53 | 146>=dif<=153 | 0.70 | 316>=dif<=328 | 0.87 |
| 26>=dif<=32 | 0.54 | 154>=dif<=162 | 0.71 | 329>=dif<=344 | 0.88 |
| 33>=dif<=39 | 0.55 | 163>=dif<=170 | 0.72 | 345>=dif<=357 | 0.89 |
| 40>=dif<=46 | 0.56 | 171>=dif<=179 | 0.73 | 358>=dif<=374 | 0.90 |
| 47>=dif<=53 | 0.57 | 180>=dif<=188 | 0.74 | 375>=dif<=391 | 0.91 |
| 54>=dif<=61 | 0.58 | 189>=dif<=197 | 0.75 | 392>=dif<=411 | 0.92 |
| 62>=dif<=68 | 0.59 | 198>=dif<=206 | 0.76 | 412>=dif<=432 | 0.93 |
| 69>=dif<=76 | 0.60 | 207>=dif<=215 | 0.77 | 433>=dif<=456 | 0.94 |
| 77>=dif<=83 | 0.61 | 216>=dif<=225 | 0.78 | 457>=dif<=484 | 0.95 |
| 84>=dif<=91 | 0.62 | 226>=dif<=235 | 0.79 | 485>=dif<=517 | 0.96 |
| 92>=dif<=98 | 0.63 | 236>=dif<=245 | 0.80 | 518>=dif<=559 | 0.97 |
| 99>=dif<=106 | 0.64 | 246>=dif<=256 | 0.81 | 560>=dif<=619 | 0.98 |
| 107>=dif<=113 | 0.65 | 257>=dif<=267 | 0.82 | 620>=dif<=735 | 0.99 |
| 114>=dif<=121 | 0.66 | 268>=dif<=278 | 0.83 | dif>=736 | 1.00 |

Source: Albers and Vries (2001)

The win expectation is presented as an illustrative example of the method that will be used in creating the informational asymmetry index. It is the object of this study to determine the proper probability distribution according to the differences in determining rankings and a suitable constant k. For smaller k values, the rating is too slow to change, and so the rating will not properly measure the perception of informational asymmetry at a determined moment.

For large k values, the rating is too sensitive a perception of recent analyst opinions. Sonas (2002) analyzed 266,000 chess games between 1994 and 2001 using different k-factor values to determine how accurate the ratings were at predicting future results, and concluded that 24 is the most accurate k-factor value, as shown in Figure 1.

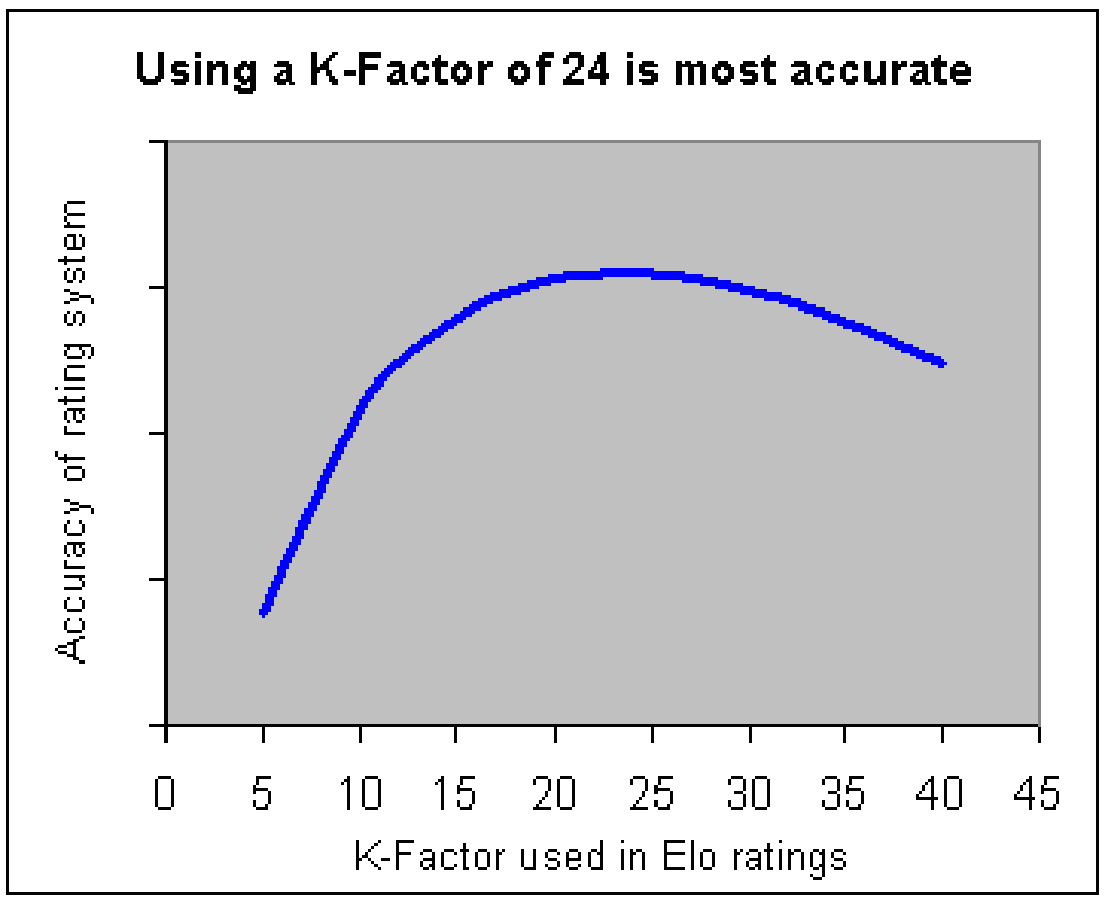

**Figure 1 – Sonas (2002) k-Factor accuracy**

In order to verify what Sonas (2002) had found, this work will test the following k-factors: 16, 24, 36, 64 and 80. In addiction the results are going to be compared among the different collected date (October, 2016 and March, 2017) to verify its stability. Also, the Sperman correlation among dates and different constant value will be tested. The following formula presents the usage of Sperman's rho:

Equation 2: Spermans's rho

$$\rho_s = 1 - \frac{6\sum_i^n d_i^2}{n(n^2-1)}$$

Source: The Author

where:

$n$ = number of firms and $d_i$ = difference between the ranks of alternative firm in the pair of rankings compared.

The k-factor value used to calculate the disclosure index was 24, the value which results in the highest Spearman rank-order correlation coefficient and is consistent with the value that Sonas (2002) considers to be most accurate.

**Data collection**

The authors have developed a website (http://www.disclosureindex.com/br), on which the current project is presented and analysts are requested to state whether or not they are certified, and their state of residence. After the analysts have completed this simple form and sent the information requested, the site presents two companies, of which the analysts choose the one they consider to have better information, i.e., where there is less information asymmetry between the company and the market. Ten pairs of firms were presented each time; some analysts participated more than once, in which case they were presented with another ten pairs of firms.

For this research, all firms composing the Brazil Broad-Based Index were included. These are the firms with stocks actively traded and analyst coverage. In October, 2016, they numbered 116, two of which had returned to being privately owned by March, 2017, when four new firms were included in the index, thus totaling 118 firms, of which 114 participated in both samples. In December, 2018 was included more 2 firms, and the index was composed by 120 firms and the fourth and last index was collected in December, 2018 with 126 firms.

This study was sponsored by three analyst associations: the CFA Institute; APIMEC (Association of Capital Market Analysts and Investment Professionals), and AMEC (Association of Investors in the Capital Markets), which invited their associates to participate by sending them an email.

The first partial rating was built on October 6, 2016, after 41 analysts had chosen the firm with the best disclosure from 712 pairs of firms. The second partial rating was built on March 10, 2017 after 52 analysts had chosen the firm with best disclosure from 932 pairs of firms. The third rating was built on December 10, 2017 with contribution of 37 analysts and the dispute of 578 pairs of companies. The Fourth and last rating was built on December 14, 2018 with the participation of 38 analysts who made their choices in 556 pairs.

**Independent Variable**

Some researchers approach that the true asymmetry value is known by the company, since they possessed all the information, also determine what will be disclosed to the market. However, the level of information asymmetry varies hugely among companies, even the ones who follow the same disclosure and corporate governance protocol. Thus, manager haven’t full control of the information asymmetry value of their own companies. It might take place since they might not be awarded over the market interest and if the investors truly understand the information disclosed by the company.

Verrechia (1983) stated that there is an equilibrium of asymmetry information which is carefully decided by managers and the market. Managers have incentive to withhold information, especially the negatives ones, and traders are aware of it, until a certain limit. The absence of information might lead investors to misprice the company's stocks, which is not the goal of any manager.

It is a possible situation that a company intend to be transparent, adopting the best practices of corporate governance, but instead of decreasing the information asymmetry, it increases because investors don't perceive its transparency. Although may be true a situation which investors are believing in company transparency, despite the company aren't adopting a full disclosure policy to the market. This last situation might be rapidly corrected by the index.

Hence, the value of information asymmetry is to sensitive over the analyst's perceptions of company disclosure, not even in the present days, but about future guidance too. Corporate communication goes from managers to intermediates, investors and savers. It can take place by different sources, directly through financial reports, press releases and media, or indirectly through financial intermediaries and financial analysts. Despite it, one of the mains roles of corporate disclosure is to eliminate agency problems (Healy & Palepu, 2001). For this reason, this works aims to understand how the analysts build their perceptions and which proxies are the most relevant to capture the value of company information asymmetry.

Literature has brought so far, an extensive enhance to proxy information asymmetry, as it can be perceived over different formats and degrees impacting cost of equity, pricing, stock price volatility and others. In order to prove the IAI consistence this paper will follow Clarke and Shastri (2001) proxy segregation, diving in three groups: external analysis, internal analysis and market microstructure.

**External Analysis**

Analyst of financial market use information provided by the company to make forecast about firm's future reports. Theses information used to come over financial reports (quarterly and annual), investors relation events and other forms of firm's communication, in additional a great analyst would study company's industry and competitors. Hence, analysts' perceptions and recommendation (buy, hold or sell stocks) are an important source of information for investor. Healy and Palepu (2001) found indicatives that analysts forecast, and recommendations add value to capital market, companies with greater coverage rapidly adjust their stock price due to new information. Although there are evidences that analyst forecast affect stock price if they are bias. In Brazilian market, sell side usually issue companies' figures individually and industry, stock price recommendation and future results.

We follow Shawn (2002) in our external analysis proxies for two reasons. The analyst forecast is verified the month before actual earnings release, by proxying in this short term the optimisms bias is avoided, consistent with Brown et al. (1985). In addition, errors in forecast made very close to earnings announcement are associated with firms-specific information rather than economy, or industry miss information.

The first measure is ERROR which is simple the absolute difference between the actual earnings and the Bloomberg median forecast deflated by the stock price five days before earnings announcement date. As literature suggested, higher differences are attributed to companies with higher degree of information asymmetry, hence is expected that those companies appear with low score in the IAI.

The second measure is COVERAGE, is the number of sell side analyst on Bloomberg data base covering the company. Lang and Lundholm (1993) found that companies with best disclosure practices have larger analyst following, as well as less analyst forecast dispersion and less volatility. Consequently, is expected that higher number of analysts following the firm, lower would be the information asymmetry, as lower would be the IAI score.

| Proxy | Formula | Expected Result | Reference |
|---|---|---|---|
| ERROR | $AE - {}^{mdFore}/_{5\ bu}$ | (-) | Shawn (2002) |
| COVERAGE | #analyst | (+) | Lang and Lundholm (1993) |

**Internal Analysis**

An extensive group of researchers dedicated to study information asymmetry on companies through their activities in the capital market, i.e. stock issue, debt issue. Its moment is particularly important because firms engage in roadshow and investor conference to increase voluntary disclosure and private channel communication targeting analyst and investors with publicly available presentation (Schiemann et al., 2010). Focusing in amplifying transparency, companies aim to decrease opacity, hence decreasing cost of capital, bid-ask spreads and increasing market liquidity (Diamond and Verrechia, 1991).

The bid-ask spread (BaA) was used as a measure of information asymmetry in most research projects, namely Chung (2006), Kanagaretnam, Lobo, and Whalen (2007), Chen, Chung, Lee, and Liao (2007), Wang and Zhang (2009), Chu and Song (2010), and Fauver and Naranjo (2010).

The rationale of using the bid-ask spread can be obtained from Glosten and Milgrom (1985), who consider that argument spreads are consequences of asymmetric information among market participants.

On the other hand, Huang and Stoll (1997) find that the bid-ask spread can be broken down into the cost of processing orders, carrying costs, and the cost of adverse selection. However, according to the authors, the most important part in determining the bid-ask spread is the cost of processing.

Moreover, the intuition of using the bid-ask spread as a proxy to measure the asymmetry of information comes from the concept of Diamond and Verrecchia (1991), in which asymmetric information reduces the liquidity of the share. The bid-ask spread can be used as a measure of liquidity of an action, and it would also be a measure of information asymmetry; however, the fact that information asymmetry decreases the liquidity of a share is not the only factor that impacts liquidity and, consequently, the bid-ask spread.

Dierkens (1991) studied the importance of information asymmetry for firms during the process of equity issuance. The paper defined information asymmetry as a determination by assets' characteristic and manager and market behavioral. By proxy information asymmetry surround equity issue, she used the standard deviation of the daily stock price abnormal return for the subsequent year of issuance, the ratio of numbers of outstanding shares traded before and after the issuance, a dummy for public announcements and for growth opportunities proxy the ratio of market value of the equity and the book value of the equity.

Adapting Dierkens (1991) proxies, VOL will be use as a proxy for asymmetry information measured by the ratio of standard deviation of daily stock price variation of firm $i$ and the standard deviation of daily variation of the Brazil Broad-Based Index (IBrA). This proxy can be associated with the number of uninformed traders presented in firm, as suggested by Wang (1993), the greater is the percentage of uninformed traders, as greater will be the stock price volatility. It is expected that companies with higher levels of volatility will have higher level of information asymmetry, also lower score on IAI ranking.

Also, GO will be a measure of growth opportunities given by Tobin q, as suggested by Smith and Watts (1992), McLaughlin et al. (1998). Besides leverage has impacts on market-to-book measure, Penman (1996) argue that market can interpret higher level of leverage as risk factor which has impacts in market value. Literature suggests other problem with this proxy, the accounting data is quarterly bases, and higher levels of Tobin q can be associated with monopoly power, not growth opportunities. Although, the different base among market and accountability information, it still can measure the presence of opacity attributed to the discounted required by the investor to acquires firms' stock. This work particularly disagree with this last concern besides monopoly has obvious advantages to the company in terms of market value if investors are willing to pay a premium to its advantage, the price would be higher as to not compensate for the expected return. It is expected the higher levels of Tobin q ratio is associated with higher levels of information asymmetry and lower IAI score.

| Proxy | Formula | Expected Result | Reference |
|---|---|---|---|
| VOL | $\sigma Stock.Price$ | (-) | Dierkens (1991) |
| BaA | $\frac{Ask\ price - Bid\ price}{Price}$ | (-) | Diamond and Verrecchia (1991) |
| Tobin q | $\frac{market\ value\ of\ equity + total\ liabilities}{Total\ assets}$ | (-) | Smith and Watts (1992) |

**Market Microstructure**

The presence of information asymmetry is directly correlated with the presence of private information on the market. If there are investors more informed than other prices and stock return will be critically determined by its presence. By modelling stock liquidity and the frequency of bid and ask spread is plausible to make inferences about the likelihood of informed trading.

Easley and O'Hara (1987) develop a model which consist to measure the impact in terms of price and size order of information asymmetry. Although they'd already said that the price-trade size relationship isn't determine exclusive by information effects, it is impacted by asymmetry issue. Their model identifies that in certain market conditions the informed traders would trade solely large trades sizes, hence small traders are uncapable to determine the asset price. They also noticed that market makers do not know if they are dealing with informed, or uninformed traded, furthermore, there is always an uncertainty if in fact a new information exists. As a result, there is a partial price equilibrium in which a large number of informed traders have small effect on the price. They also noted that informed traders maximize the expected profit of each trade individually, not in aggregate terms, what the authors called the competitive behavior.

The existence of uninformed investors is the roots of information asymmetry and a plausible reason to market imperfection. Wang (1993) exploited a dynamic model of asset pricing under asymmetric information, identifying that uniformed traders contributes with market volatility. Investors are concerned about future cash flow, and noise traders, to determine stock price, when investors are less informed about company expectation of dividends growth rate, it becomes a harder task. In order to diminished noise traders, investors demand higher premium, turning prices more elastic to supply shocks. There is a positive relation between the existence of uninformed traders and higher premia, as information became less spread, stock price will not reflect companies' fundamentals, increasing risk premia.

Because literature shows that traders demand higher return to invest in companies which have greater private information, Easley, Kiefer and O'Hara (1997), Easley, Hvidkjaer, and O'Hara (2002), Easley and O'Hara (2004), develop an equilibrium model to price information asymmetry, shading light in the assumption that well informed investors play a critical component on price formation process (Hvidkjaer, 2010). Since informed investors are capable the rebalance their portfolios when news information arrives, uniformed investors are always on the wrong side, holding to many stocks with bad news and a few with good news, in a frustrated attempt to diminish the risk by diversification. Hence, the presence of private information in the market shows that the CAPM theory is wrong about systematic risk, investors won't hold identical portfolios, because the expected return and risk perception aren't the same among then, uniformed investors requires a greater return to hold stock and are more sensitive to new information, on the other hand because informed investors know which stocks have good and bad news, they are able to hold (or even sell bad stocks) for a longer time.

As a proxy for information asymmetry, Easley, Engle, O'Hara y Wu (2008) continue the studies and develop a dynamic model measuring the interaction among informed and uniformed investor in terms of liquidity, market depth and order flow through time. Literature uses the probability of informed trading (PIN) as one of most accurate proxies to measure asymmetry. It is based on the theoretical assumption that informed trader are the ones who unbalance the trade equilibrium, by using their private information it's possible to infer that abnormal return are plausible, since they are always on the right side of the trade, causing the adverse selection problem to uniformed investors.

This work will follow the contributions of Ealey and O'Hara and Hvidkjaer model to determine the probability of informed trading. Hvidkjaer (2010) suggested a tree diagram of trading process which good news (δ) or bad news (1-δ) occur with α probability at a date $t$, changing the stock price to $\bar{V}_t$ if there is good news arriving, or to $V_t$ if it is bad news, as suggested in the chart below.

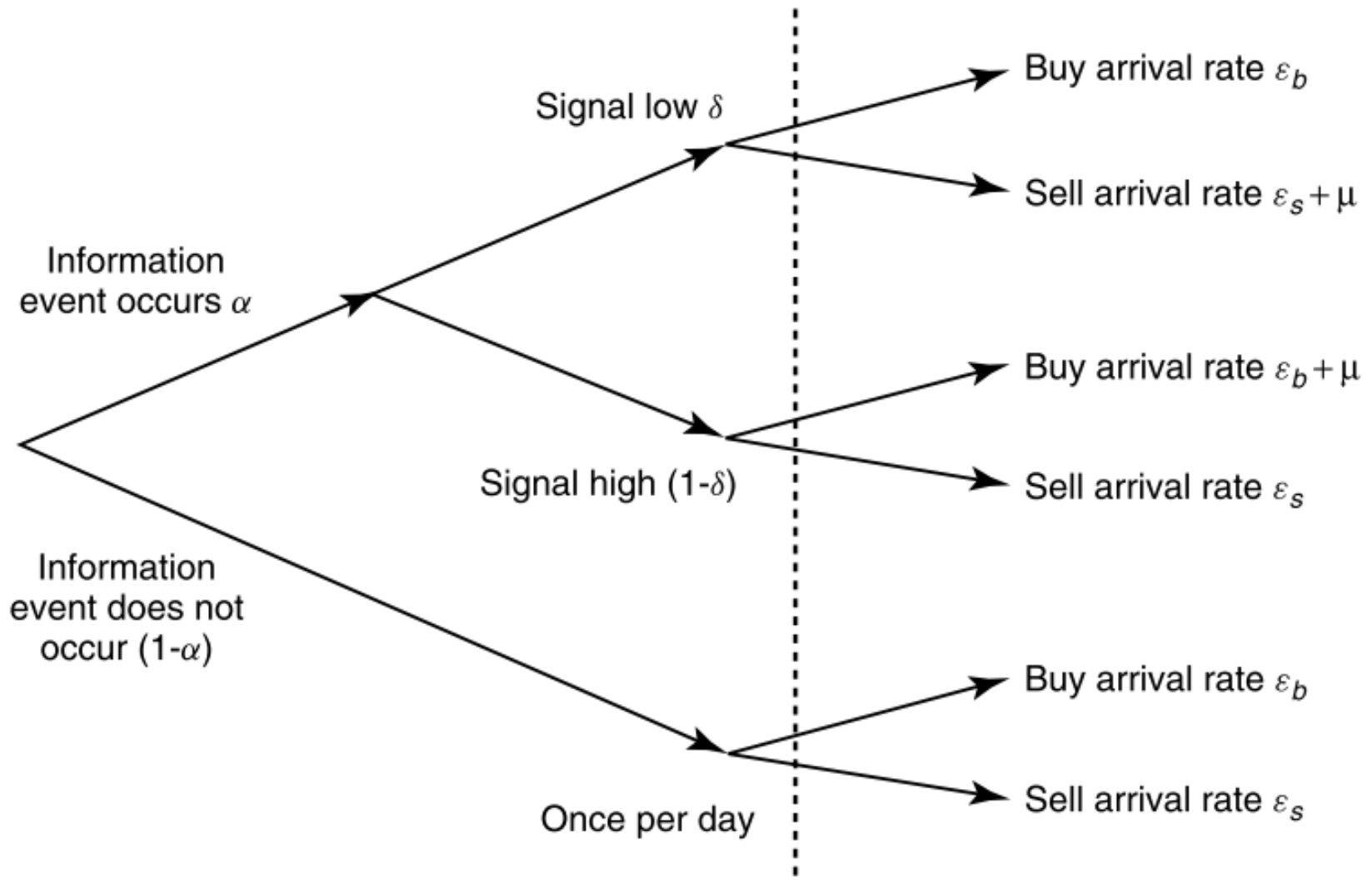

During a trading day, investors place their orders according a Poisson process executing then according their own necessity. Informed investors arrive at rate μ as uniformed investors – $\varepsilon_b$ for buyers and $\varepsilon_s$ for sellers – trades for liquidity reasons.

Estimating via maximum likelihood is possible to determine the PIN of stock $j$ at date $t$. With the help of Bloomberg platform, the number of buyers and sellers of a day give the first step of the estimation. The follow equation is the likelihood formed by these investors, where $B$ is the total number of buyers, $S$ is the total number of sellers and $\theta$ $\theta = (\mu, \varepsilon_b, \varepsilon_s, \alpha, \delta)$ is the parameter vector. As suggested by the diagram, this likelihood function is weighted by the probability of good news takes place, bad news takes place, or even no news at the date.

Equation 3: PIN

$$\mathcal{L}\left((B,S)\ \middle|\theta\right) = \alpha(1-\delta)e^{-(\mu+\varepsilon_b+\varepsilon_s)}\frac{(\mu+\varepsilon_b)^B(\varepsilon_s)^S}{B!\,S!}$$
$$+\alpha\delta e^{-(\mu+\varepsilon_b+\varepsilon_s)}\frac{(\mu+\varepsilon_b)^B(\varepsilon_s)^S}{B!\,S!}$$
$$+(1-\alpha)e^{-(\varepsilon_b+\varepsilon_s)}\frac{(\varepsilon_b)^B(\varepsilon_s)^S}{B!\,S!}$$

Source: The Author

Following Hvidkjaer (2010) in order to increase computing efficiency and reduces truncation error the likelihood function as rearranged to the following equation, where $M_t = min(B_t, S_t) + \max(B_t, S_t)/2$, $x_s = \varepsilon_S/\mu + \varepsilon_S$, $x_b = \varepsilon_b/\mu + \varepsilon_b$,

Equation 4: $\ln$ PIN

$$\mathcal{L}\left((B_t, S_t)_{t=1}^{T}\ \middle|\theta\right)$$
$$= \sum_{t=1}^{T}\left[-\varepsilon_b - \varepsilon_s + M_t(\ln x_s) + B_t\ln(\mu+\varepsilon_b) + S_t\ln(\mu+\varepsilon_{s)}\right]$$
$$+\sum_{t=1}^{T}\ln\left[\alpha(1-\delta)e^{-\mu}x_S^{S_t-M_t}x_b^{-M_t} + \alpha\delta e^{-\mu}x_S^{-M_t} + (1-\alpha)x_S^{S_t-M_t}x_b^{B_t-M_t}\right]$$

Source: The Author

The PIN estimation can overpost the conceptual of bid and ask spreads as they came from the same theory, nevertheless, is more robust Easley et al (2002). In addition PIN has an extensive impact on companies: higher cost of capital Duarte, J., Han, X., Harford, J. & Young, L. (2008), presence of insider trading Aslan, H., Easley,

D., Hvidkjaer, S. & O'Hara, M. (2007), higher expected returns Easley, D. & O'Hara, M. (2004), suggesting that higher PIN higher information asymmetry. Hence, is expected that higher levels of PIN would be associated of score o IAI ranking.

| Proxy | Formula | Expected Result | Reference |
|---|---|---|---|
| PIN | Equation 4 | (-) | Hvidkjaer (2010) |

**Econometric Model**

This work aims to build a high accuracy model in order estimate the companies' disclosure from the perspective of the market analyst. In this sense, care is taken to select the right and best-fitted model.

We include also 3 control variables: size (total assets), free float and volume.

As the dependent variable will be run for Three periods, the regression model would be a panel with pooled cross-section data, as follow:

Equation 5:

$$IAI = \alpha + \beta_1 Cov + \beta_2 Error + \beta_3 Vol + \beta_4 qtobin + \beta_5 BaA + \beta_6 PIN + \beta_7 lnsize + \beta_8 ff + \beta_9 lnvolume + \mathcal{E}$$

Source: The author

The literature review stated that the market analyst's perception could change over time. Facing that, a time dummy would be introduced for every period. Additionally, sectors dummies also are going to be introduced trying to avoid that specifics risk of a sector influence companies from different sectors. The Hausman test is going to be applied to verify if the effects are random or fixed what has impact in the regression. Also, robustness test will be perform to emphasize the IAI's efficacy.

## 4 RESULTS

The Descriptive statistics are presented in Table 6. We have 342 observations from 3 ratings. The first rating was not considered because the Index was not well consolidated.

The variable errors in forecast has many missing data and have only 174 observations. The first variable is our dependent variable, that is our ranking according analyst perception and it varies between 1315 and 1731.

**Table 6: Descriptive statistics**

| Variable | Obs | Mean | Std. Dev. | Min | Max |
|---|---|---|---|---|---|
| ranking | 342 | 1505.064 | 80.72462 | 1315 | 1731 |
| Coverage | 342 | 10.55263 | 4.799973 | 0 | 21 |
| Error | 174 | -.2411871 | .7308159 | -3.2585 | 3.5986 |
| Vol | 342 | 40.67281 | 19.03582 | 1.67 | 145.593 |
| BaA | 342 | .2063833 | .2282664 | .0444 | 2.673 |
| lnvolume | 342 | 16.24797 | 2.100477 | 9.277999 | 20.86594 |
| lnsize | 342 | 23.26457 | 1.565775 | 19.10888 | 28.05268 |
| ff | 342 | 58.24518 | 25.95821 | 4.05 | 100 |
| qtobin | 339 | 1.572493 | 1.142273 | .5829 | 11.2332 |

The correlation between the variables are shown in Table 7. As can be observed, none of the variables present a high correlation.

**Table 7: Correlation Matrix**

| | Coverage | Error | Vol | qtobin | BaA | lnvolume | lnta | ff |
|---|---|---|---|---|---|---|---|---|
| Coverage | 1.0000 | | | | | | | |
| Error | 0.2286 | 1.0000 | | | | | | |
| Vol | -0.1402 | -0.1192 | 1.0000 | | | | | |
| qtobin | 0.0224 | 0.0650 | -0.0316 | 1.0000 | | | | |
| BaA | -0.4138 | 0.0378 | 0.2763 | -0.0951 | 1.0000 | | | |
| lnvolume | 0.2731 | -0.1316 | -0.1056 | 0.0996 | -0.3727 | 1.0000 | | |
| lnta | 0.1623 | -0.0229 | -0.1262 | -0.3494 | -0.3445 | 0.3658 | 1.0000 | |
| ff | 0.1827 | -0.0311 | 0.0027 | -0.1705 | -0.1637 | 0.0610 | -0.1808 | 1.0000 |

The highest correlation is between Coverage and Bid-Ask spread (BaA). It is expected that firms with lower coverage present higher Bid-Ask spread. Since it was not necessary to remove any variables, the regression model was run. The results are presented in Table 8.

**Table 8: Regression model**

| Source | SS | df | MS |
|---|---|---|---|
| Model | 386507.536 | 8 | 48313.442 |
| Residual | 854821.585 | 164 | 5212.32674 |
| Total | 1241329.12 | 172 | 7217.02978 |

| | |
|---|---|
| Number of obs = | 173 |
| F( 8, 164) = | 9.27 |
| Prob > F = | 0.0000 |
| R-squared = | 0.3114 |
| Adj R-squared = | 0.2778 |
| Root MSE = | 72.196 |

| ranking | Coef. | Std. Err. | t | P>\|t\| | [95% Conf. | Interval] |
|---|---|---|---|---|---|---|
| Coverage | 1.733812 | 1.338451 | 1.30 | 0.197 | -.9090061 | 4.37663 |
| Error | -7.955356 | 8.111772 | -0.98 | 0.328 | -23.97233 | 8.061618 |
| Vol | -1.037437 | .3450925 | -3.01 | 0.003 | -1.718834 | -.3560403 |
| BaA | 1.896637 | 47.35702 | 0.04 | 0.968 | -91.61144 | 95.40471 |
| lnvolume | 4.566208 | 3.150041 | 1.45 | 0.149 | -1.653656 | 10.78607 |
| lnsize | 19.69807 | 4.307993 | 4.57 | 0.000 | 11.19179 | 28.20435 |
| ff | -.031324 | .2295619 | -0.14 | 0.892 | -.4846019 | .421954 |
| qtobin | 17.37489 | 4.988945 | 3.48 | 0.001 | 7.52405 | 27.22574 |
| _cons | 976.4072 | 106.735 | 9.15 | 0.000 | 765.6552 | 1187.159 |

The model presented a good fit between our index and the proxies used to measure informational asymmetry. The F value is significant and the $R^2$ is 27.78%, what we consider suitable for the proposed model. In this first stage we found 3 of the 8 variables as significant.

We extracted the least significant variables one by one from the model and arrived at a final model with 4 variables: coverage, volatility, Tobin q and size. The regression model is presented in Table 9.

**Table 9: Final regression model**

| Source | SS | df | MS |
|---|---|---|---|
| Model | 628577.5 | 4 | 157144.375 |
| Residual | 1579557.04 | 334 | 4729.21269 |
| Total | 2208134.54 | 338 | 6532.94242 |

| | |
|---|---|
| Number of obs | = 339 |
| F( 4, 334) | = 33.23 |
| Prob > F | = 0.0000 |
| R-squared | = 0.2847 |
| Adj R-squared | = 0.2761 |
| Root MSE | = 68.769 |

| ranking | Coef. | Std. Err. | t | P>\|t\| | [95% Conf. | Interval] |
|---|---|---|---|---|---|---|
| Coverage | 2.23952 | .8365905 | 2.68 | 0.008 | .5938692 | 3.88517 |
| Vol | -.8197953 | .198787 | -4.12 | 0.000 | -1.210828 | -.4287629 |
| lnsize | 20.01405 | 2.65614 | 7.54 | 0.000 | 14.78918 | 25.23893 |
| qtobin | 23.09268 | 3.529336 | 6.54 | 0.000 | 16.15015 | 30.03521 |
| _cons | 1012.343 | 62.89146 | 16.10 | 0.000 | 888.6301 | 1136.057 |

The final model with only 4 variables presented almost the same $R^2$ than the model with 8 variables, but with lower Root MSE indicating better fit.

The coverage, as expected contribute for reduce the information asymmetry. Obviously, the more analysts monitoring and analyzing the company, the less informational asymmetry will be, as already identified by Lang and Lundholm (1993).

Wang (1993) believed that the greater is the percentage of uninformed traders, as greater will be the stock price volatility and this expectation was confirmed by our model. The coefficient of the stock price volatility is negative.

Size is a control variable, and our intuition is that the larger firms should present lower informational asymmetry and this was confirmed by our model.

The last variable from the model is the Tobin q that was a proxy for growth opportunity as suggested by Smith and Watts (1992). It was expected that firms with higher growth opportunity have more strategic information and projects unknown by the market and consequently more informational asymmetry. However, contrary to that expected, the coefficient signal of this variable showed a positive sign.

This result can be interpreted as a reverse causality. Tobin's q is rather associated with greater growth opportunities, but Tobin's q is perhaps primarily a measure of value.

In this sense, it would be expected that companies with lower levels of informational asymmetry would be valued by the market and consequently would have a greater Q of Tobin.

Our interpretation then is that the positive coefficient of Tobin's q is associated with the valuation that the market gives to the best informational level presented by the company.

## 5 CONCLUSION

The purpose of the research was to build an index of informational asymmetry with market and firm proxies that reflect the analysts' perception of the level of informational asymmetry of companies. This paper adds to disclosure research since it provides a reliable proxy to be used in the field. Several studies have aimed to find the benefits of good disclosure for firms, but they have as a limitation the absence of a good proxy for the disclosure level of a company. This study considers that the perception of analysts is a good proxy for the disclosure level of a company, and a tool was constructed to collect analysts' perception and create an index with an algorithm based on Elo ratings.

After four waves of analyst participation, four rankings were calculated. The high correlation found between the different rankings eliminates the possibility that the answers were random, and confirms that the disclosure index described in this paper indicates analysts' perception, and is a good proxy for use as a measure of disclosure of Brazilian firms.

Obtaining the participation of analysts is an arduous and very costly task. Therefore the ultimate goal was to create a model with market and company variables that reflect analysts' perceptions.

Our model presented an excellent adjustment between the index constructed through the analysts' perception and variables used in the literature as possible proxies for informational asymmetry, which reinforces the validity of our index.

The variable coverage, volatility and size presented the expected relation with our informational asymmetry index, however, the Tobin q, used as a proxy for growth opportunity presented positive coefficient, contrary to our expectations. Our interpretation for this result is that it is capturing the valuation of the company by presenting a lower level of information asymmetry.

In any case, our objective has been reached and the model presented can be used for other researches that need to measure the level of informational asymmetry of a company.